\pdfoutput=1

\documentclass[aps,prl,twocolumn,showpacs,amsmath,amssymb,amsfonts,long,superscriptaddress,preprintnumbers]{revtex4}
\usepackage{graphicx,color,soul}

\newcounter{mysectionnumber}
\newcommand{\mysection}[2]{\addtocounter{mysectionnumber}{1}\textbf{\arabic{mysectionnumber}.} \textsl{#1}{#2} -- }

\begin{document}

\title{Short-baseline neutrino oscillations, Planck, and  IceCube}

\author{John F. Cherry}
\affiliation{Center for Neutrino Physics, Department of Physics, Virginia Tech, Blacksburg, VA 24061, USA}
\author{Alexander Friedland}
\affiliation{SLAC, Stanford University, Menlo Park, California 94025 USA}
\email{alexfr@slac.stanford.edu}
\author{Ian M. Shoemaker}
\affiliation{Department of Physics, The Pennsylvania State University, University Park, PA 16802, USA}
\email{shoemaker@psu.edu}
\date{May 20, 2016}

\begin{abstract}
We examine a framework with light new physics, which couples to the Standard Model only via neutrino mixing. Taking the hints from the short-baseline anomalies seriously and combining them with modern cosmological data and recent IceCube measurements, we obtain surprisingly effective constraints on the hidden force: keV $\lesssim M \lesssim0.3$ GeV for the mediator mass and $g_{h}>10^{-6}-10^{-3}$ for the coupling constant. Flavor equilibration between the hidden and active neutrinos can be delayed until temperatures of $\sim 1$ MeV, but not below $\sim 100$ keV. This scenario can be tested with next-generation Cosmic Microwave Background,  IceCube, and oscillation experiments.

\end{abstract}

\preprint{SLAC-PUB-16517}

\maketitle

\mysection{Introduction and motivation} {.} 
There has been a growing interest in the existence of \emph{secluded} sectors containing new light fields and hidden forces.
To escape detection to date, such sectors must couple to the Standard Model (SM)  extremely weakly and the nature of this coupling (\emph{portal}) is the 
crucial issue dictating search strategies. A considerable amount of recent effort has been concentrated on the kinetic mixing (``dark photon'') portal ({\it e.g.}, \cite{Essig:2013lka}). In contrast, \emph{the neutrino mixing portal} is not yet sufficiently explored. Yet, it is also physically well motivated and, moreover, may have even been observed in neutrino oscillation experiments.

At first, it might seem that it would be difficult to learn much about the properties of the dark sector in this case. This is, however, not so. As shown here, by combining diverse experimental input -- from terrestrial, astrophysical and cosmological neutrino sources -- it is possible to detect the presence of the hidden force in the entire allowed parameter space. In the process, we revisit the question whether hidden neutrino interactions can reconcile short-baseline oscillation hints with modern cosmological constraints. Our findings differ from those of several recent papers on the topic~\cite{Hannestad:2013ana,Dasgupta:2013zpn,Bringmann:2013vra,Archidiacono:2014uo,Saviano:2014esa,Mirizzi:2014ama,Chu:2015ipa}.

\mysection{Theoretical framework}{.} The setup is the same as previously used in \cite{Cherry:2014xra}. The secluded sector contains a light fermion, $\nu_{h}$, which is a SM singlet, but carries hidden interactions mediated by light (but not massless) gauge bosons. For simplicity, we take the hidden gauge group to be $U(1)_{h}$, though the argument  applies to general groups.  At dimension 5, in addition to the standard Weinberg operators $(LH)(LH)/\Lambda$ and  $(\nu_{h} \eta)(\nu_{h} \eta)/\Lambda$, one also has the cross terms $(LH)(\nu_{h} \eta)/\Lambda$. Here $L$ are the SM lepton doublets containing active neutrinos $\nu_{a}$, $H$ is the SM Higgs doublet, $\eta$ is the secluded Higgs field responsible for giving the hidden vector mediator $A'$ its mass, and $\Lambda$ a high energy scale. The first two operators give small Majorana masses to $\nu_{a}$ and $\nu_{h}$ (thus justifying the assumption of lightness), while the last one mixes $\nu_{h}$ and $\nu_{a}$. A simple realization involves heavy right-handed singlet fermions, which provide the seesaw mechanism in the hidden and SM sectors and link the two. Such constructions have been known for many years, for example in the context of the ``mirror worlds'' ({\it cf}. \cite{Berezhiani:1995yi}). 

\mysection{Experimental hints}{?}  
Over 20 years ago, the LSND experiment reported \cite{Athanassopoulos:1996jb,Athanassopoulos:1996wc,Louis:1995vg} what was believed to be  evidence for short-distance $\bar\nu_{\mu}\rightarrow\bar\nu_{e}$ flavor conversion.
This result, which cannot be explained within the 3-flavor paradigm, has both excited and confounded the community ever since. Despite extensive follow-up efforts by experiments such as MiniBOONE \cite{Aguilar-Arevalo:2013pmq}, this claim has not been conclusively ruled out. Meanwhile, cosmology entered the era of precision measurements. Recent results from the Planck collaboration~\cite{Ade:2015xua} provide impressive constraints on the number of effective neutrino species at the CMB formation epoch: $N_{eff} = 3.15 \pm 0.23$. Does this disfavor the oscillation explanation for LSND? 

\mysection{Nonsterile ``sterile'' neutrinos}{?}
The answer to the last question is known to be positive if the additional neutrinos are truly sterile (TS). Such neutrinos equilibrate with the active species $\nu_{a}$ while the latter are still coupled to the  $e^{\pm}\gamma$ bath (see below). Thus, a common temperature is established for $\nu_{s}$, $\nu_{a}$,   $e^{\pm}$ and $\gamma$, in conflict with the $N_{eff}$ measurement. 
Does this conclusion hold if the secluded neutrinos carry hidden interactions? Can one, for some choice of parameters, keep $\nu_{a}$ and $\nu_{h}$ out of equilibrium until after $\nu_{a}\bar\nu_{a}\leftrightarrow e^{+}e^{-}$ freezes out? 

\begin{figure}[htb]
  \includegraphics[angle=0,width=0.47\textwidth]{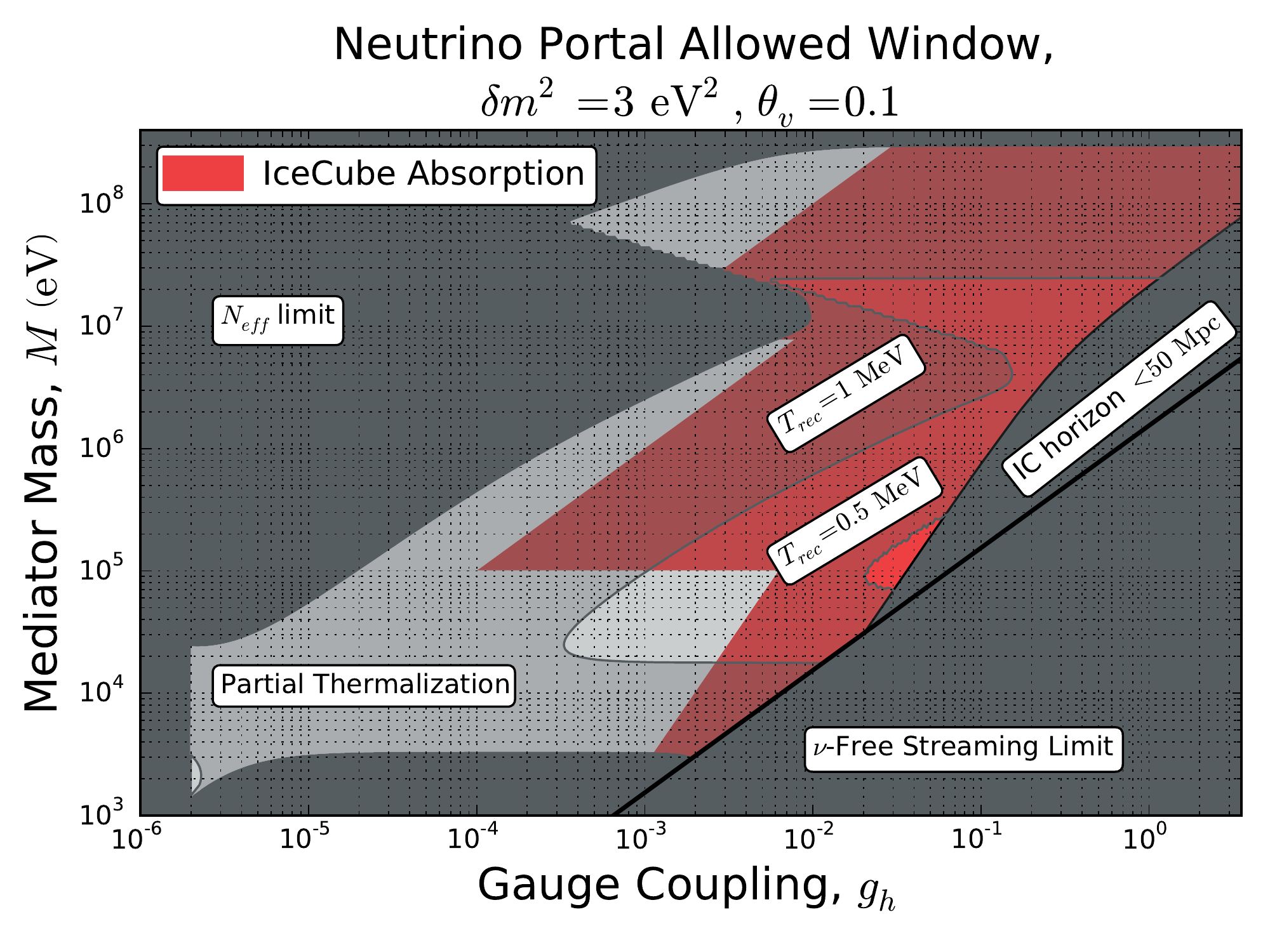}
  \caption{The allowed region and future coverage of the neutrino portal parameter space.} 
  \label{fig:allowedwindow}
\end{figure}

\mysection{Results: the secluded window}{.}
We have investigated this issue quantitatively. Fig.~\ref{fig:allowedwindow} summarizes our findings, in the space of the hidden mediator mass $M$ and coupling $g_{h}$, for a representative choice $\Delta m^{2} \sim 3$ eV$^{2}$, $\theta\sim0.1$ \cite{Aguilar-Arevalo:2013pmq,Antonello:2013gut,TheIceCube:2016oqi}\footnote{In the absence of a conclusive measurement, the oscillation parameter space is covered by a number of difficult experiments with different systematics and possible unknown errors. Our analytical expressions clearly show how the results scale with the choice of the oscillation parameters.}. The dark gray regions are excluded, by conservative criteria. The remaining window contains filled contours indicating the reach of future experiments. As one can see, the coverage of these experiments is  complete.

The conservative constraints are as follows: (i) the $\nu_{a}\leftrightarrow\nu_{h}$ recouping temperature $T_{rec}<3$ MeV; (ii) at $T\sim 1$ eV (the CMB era), most of the neutrinos in the universe \emph{stream freely}, unaffected by the new $\nu-\nu$ interactions; and (iii) the mean free path for PeV astrophysical neutrinos is $>50$ Mpc.
The last ingredient implies the absence of a local horizon for these neutrinos, since they do not strongly correlate with the local structure in the IceCube data. The impact of each of these constraints is labeled in the dark-gray regions in Fig.~ \ref{fig:allowedwindow}. 
It is remarkable that such basic considerations limit the hidden force to a specific window. 
We again stress that our findings differ from those in the literature.

Now let us turn to the impact of future experiments. First, in the allowed window, the recouping $\nu_{a}\leftrightarrow\nu_{h}$ temperature lies in the interval $T_{rec}\sim0.5-3$ MeV, close to the $\nu_{a}\bar\nu_{a}\leftrightarrow e^{+}e^{-}$ freezeout.  One thus expects a fractional deviation of $N_{eff}$ from 3.04, which should be detectable by future CMB measurements ({\it e.g.}, CMB-S4 \cite{CMBS4}). Second, a large part of the window is expected to give absorption features in the astrophysical neutrino flux observable by the upgraded (10 km$^{3}$)  IceCube  (IC). The relevant sensitivities are labeled  \emph{Partial Thermalization}  and \emph{IceCube absorption}.

Next, we survey the relevant physics underlying Fig.~\ref{fig:allowedwindow}. A number of different regimes  must be considered, and there are a number of subtle effects. For clarity, we suppress numerical coefficients in our equations (our numerical calculations were carried out with all relevant factors).

\mysection{Flavor equilibration basics}{.}
The populations of $\nu_{a}$ and $\nu_{h}$ can come into equilibrium via a combined action of oscillations and collisions. 
For flavor equilibration to occur, the effective flavor change rate, $\Gamma_{fc}$, must at least equal to the expansion rate $\Gamma_{exp}\sim T^{2}/ M_{pl}$, where $M_{pl}$ is the Planck mass.
The rate $\Gamma_{fc}$, in turn, is
\begin{equation}
\label{eq:convrate}
\Gamma_{fc}=P(\nu_{a}\rightarrow\nu_{s})\Gamma_{c}=P(\nu_{a}\rightarrow\nu_{s})\sigma_{c} n, 
\end{equation} 
where $P(\nu_{a}\rightarrow\nu_{h})=\sin^{2} 2\theta\sin^{2}(\Delta m^{2}L/4E)$ is the oscillation probability between collisions, depending on the distance $L$, neutrino energy $E$, mixing angle $\theta$ and the mass-squared splitting $\Delta m^{2}$, and $\Gamma_{c}$ is the collision rate. 

The standard picture of the TS neutrino equilibration in the early universe is reviewed in the Appendix, for completeness. In brief, flavor equilibrium obtains in a window of temperatures, from
\begin{equation}
\label{eq:QZ}
 T_{rec}\sim (\sin^{2} 2\theta (\Delta m^{2})^{2} G_{F}^{-2} M_{pl})^{1/9}.
\end{equation}
where $G_{F}$ is the Fermi constant, to 
\begin{equation}
\label{eq:Tts}
 T_{dec} \sim [G_{F}^{2} M_{pl}(1/2)\sin^{2} 2\theta]^{-1/3}.
\end{equation}
Below $T_{dec}$ flavor conversion ceases because the weak interactions (WI) freeze out. Above $T_{rec}$, it is suppressed by frequent collisions \cite{Stodolsky:1986dx} (the ``Quantum Zeno'' effect). $\Gamma_{fc}$ is maximized when the frequencies of collisions and oscillations match, at
\begin{equation}
\label{eq:Toptimal}
 T_{opt}^{wi} \sim ( \Delta m^{2}/G_{F}^{2})^{1/6}.
\end{equation}
For $\Delta m^{2}\sim 3$ eV$^{2}$, this gives $T_{opt}^{wi} \sim 50$ MeV.
 
 The SM process $\nu\bar\nu\leftrightarrow e^{-}e^{+}$  freezes out at~\cite{Kolb:1990vq}
\begin{equation}
\label{eq:Tenu}
 T_{dec}^{e\nu} \sim (G_{F}^{2} M_{pl})^{-1/3} \sim 1\mbox{ MeV}.
\end{equation}
Up to a factor of the mixing angle, Eqs.~(\ref{eq:Tts}) and (\ref{eq:Tenu}) are the same. This is the root cause of difficulties for the TS explanation of the short-baseline anomalies: since the oscillation fit requires $\theta$ to be not too small, the flavor equilibrium will be achieved slightly above $T_{dec}^{e\nu}$.

\mysection{Hidden interactions: suppressed $\nu_{a}-\nu_{h}$ mixing}{.}The WI generate a second-order Notzold-Raffelt (NR) matter potential \cite{Notzold:1987ik,Raffelt:1996wa} $V_{NR}\sim-G_{F}^{2} T^{5}/\alpha$. This potential does not prevent equilibration. Indeed, at $T_{dec}$, Eq.~(\ref{eq:Tts}), $G_{F}^{2}T^{5}/\alpha \ll \Delta m^{2}/T$, so that in-matter mixing angle, $\theta_{m}$, is not suppressed there. 

A primordial population of $\nu_{h}$ creates, through the hidden interactions, its own term in the matter potential. If these interactions are stronger that WI, $\theta_{m}$ stays suppressed until lower temperatures than in the TS case~\cite{Babu:1991at}. This \emph{may} delay recouping of the two sectors, though there are several competing effects, as discussed below. The primordial populations of $\nu_{h}$, $A'$ and $\eta$ can be created during an early epoch when the SM and hidden sectors were coupled.  Alternatively, even if the primordial populations are zero, they will be generated by oscillations, until $T_{h}$ is high enough to suppress further conversions.

The exact value of $T_{h}$ is thus model-dependent, a function of the physics at high energy scales and the number of degrees of freedom in the secluded sector. Such primordial population would contribute a fractional $\delta N_{eff}$. In what follows we take the ratio $T_{h}/T_{SM}\sim{\cal O}(1)$, so that in the estimates we do not distinguish $T_{h}$ and $T_{SM}$.

Two regimes must be considered \cite{Dasgupta:2013zpn}. In the contact limit $M\gg T$, the form of $V_{m}$ is completely analogous to the NR potential, $V_{m}\sim -g^{2} E T_{h}^{4}/M^{4}$. In the opposite regime $M\ll T$, the potential instead follows Weldon's calculation for massless QED \cite{Weldon:1982bn}, $V_{m}\sim +g^{2}T_{h}^{2}/E$. The full calculation~\cite{Dasgupta:2013zpn} is analogous to the SM case \cite{Quimbay:1995jn}. 

The angle $\theta_{m}$ is suppressed when $|V_{m}|\gg \Delta m^{2}/2T$. We have $\sin2\theta_{m} \simeq (\Delta m^{2}/2T|V_{m}|)\sin2\theta$, or 
\begin{eqnarray}
\label{eq:sin2thetam}
\sin2\theta_{m} \sim 
\begin{cases}
      (\Delta m^{2}M^{4}/g^{2}T^{6})\sin2\theta & \text{(NR)},\\
      (\Delta m^{2}/g^{2}T^{2})\sin2\theta & \text{(Weldon)}.
    \end{cases}
\end{eqnarray}
Notice that the sign of $V_{m}$ implies that in the NR regime $\theta_{m}\rightarrow\pi/2$, while in the Weldon regime  $\theta_{m}\rightarrow0$. 

\mysection{Suppression of the WI collisions}{.}
For collisions between active neutrinos, the arguments leading up to Eq.~(\ref{eq:Tts}) continue to apply, with $\theta\rightarrow\theta_{m}$, which implies  $\Gamma_{fc}^{wi}\sim G_{F}^{2} T^{5} \sin^{2}2\theta_{m}$. Using Eq.~(\ref{eq:sin2thetam}), one finds in the contact limit $\Gamma_{fc}^{wi} \sim G_{F}^{2} T^{-7} (\Delta m^{2} M^{4}\sin2\theta/ g^{2})^{2}$, while in the massless limit $\Gamma_{fc}^{wi} \sim G_{F}^{2} T (\Delta m^{2} \sin2\theta/ g^{2})^{2}$. Comparing these to the expansion rate $T^{2}/M_{pl}$ we see that in both cases the most favorable recouping conditions now occur at \emph{low} temperature. Thus, if the WI collisions are out of equilibrium at $T\sim1$ MeV, they will be even more so at higher temperatures.

We should thus first of all require that the $\theta_{m}$ be suppressed at $T_{dec}$. The condition $|V_{m}|\gg \Delta m^{2}/2T$ becomes 
\begin{eqnarray}
\label{eq:suppress}
g \gg
      \sqrt{\Delta m^{2}}/ T_{dec} f(M/T_{dec})\sim 10^{-6} f(M/T_{dec}).
\end{eqnarray}
Here, $f(x)\rightarrow1$ when $x<1$ and $f(x)\rightarrow x^{2}$ for $x>1$. Once Eq.~(\ref{eq:suppress}) is fulfilled, we have $\Gamma_{fc} < \Gamma_{exp}$ at $T=T_{dec}$ and hence for all earlier $T$.  

\mysection{Effect of secluded interactions}{.}
The limits corresponding to Eq.~(\ref{eq:suppress}) are clearly seen in Fig.~\ref{fig:allowedwindow}. They, however, are just two of the many constraints that form the allowed region. This means that \emph{the suppression of the mixing angle does not guarantee delayed recoupling}.

Indeed, the hidden force, while suppressing conversions due to the WI collisions, itself mediates collisions. When $M$ is lighter that the weak scale and $g_{h}$ is not too small, most collisions at late times are due to the hidden force. A quantitative study of the competing effects of the suppressed $\theta_{m}$ and enhanced collision rates is required.

As in the TS case, the analysis depends on the ratio of the oscillation length $l_{osc}$ and the collision length $l_{col}$.

\mysection{$l_{osc}<l_{col}$}{.}
In this regime, we can write the rate of $\nu_{h}\rightarrow\nu_{a}$ flavor conversion by complete analogy with the TS case: $\Gamma_{fc}=(\sin^{2}2\theta_{m}/2)\sigma_{h} n_{h}$. Here, $\sigma_{h}$ is the cross section of the hidden interactions. The reverse rate, $\nu_{a}\rightarrow\nu_{h}$, is the same. 
One can think of scattering the mass eigenstates $\nu_{1}$ (predominantly $\nu_{a}$) on $\nu_{2}$ (predominantly $\nu_{h}$): the amplitude is suppressed by $\sin\theta_{m}\cos\theta_{m}$. 

Both the NR and Weldon regimes need to be considered.
In the first case, $\sigma_{h}\sim g^{4} E T_{h}/M^{4}$. When $M\ll T$, the answer is more subtle. The cross section, due to the $t$-channel exchange, is strongly forward-peaked (most flavor conversions occur in ``glancing blows''), and regulated by the mediator mass, $\sigma_{h}\sim g^{4}/M^{2}$. We stress that it is different from $g^{4}/T_{h}^{2}$ ({\it cf.} \cite{Archidiacono:2014uo,Chu:2015ipa}), which is the transport cross section.

With these ingredients, from $\Gamma_{fc}\sim\Gamma_{exp}$ we obtain
\begin{eqnarray}
\label{eq:Treccontact}
T_{rec} &\sim& T_{0}^{5/9}M^{{4/9}}, \;\;\;\mbox\;\;\; T_{rec}\ll M, \\
\label{eq:Trecmassless}
T_{rec} &\sim& T_{0}^{5/3}M^{-2/3}, \;\;\;\mbox\;\;\;T_{rec}\gg M. 
\end{eqnarray}
Here a special temperature, $T_{0}$, was introduced:
\begin{eqnarray}
\label{eq:T0}
T_{0}&=&[(\Delta m^{2}\sin2\theta_{v})^{2} M_{pl}]^{1/5}  \nonumber\\
&\sim& 10^{5} \mbox{ eV} (\Delta m^{2}/1\mbox{ eV}^{2} )^{2/5} (\sin 2\theta_{v}/0.1)^{2/5}.
\end{eqnarray}

Notice that Eqs.~(\ref{eq:Treccontact},\ref{eq:Trecmassless}) do not contain the coupling constant $g_{h}$.  Its effect in suppressing $\theta_{m}$ is canceled  by the cross section dependence. This perfectly illustrates the competing effects of the hidden interactions mentioned above. The same competition is seen in the dependence of the recoupling temperature on $M$. Even though $\theta_{m}$ becomes more suppressed as $M$ decreases, $T_{rec}$ turns around below $M=T_{0}$, as can be seen in Fig.~\ref{fig:Tdec} where numerically computed $T_{rec}$ is depicted. 

Requiring  $T_{dec}<T_{dec}^{e\nu}$, we obtain limits on the  range of the mediator masses, keV $\lesssim M \lesssim400$ MeV. These are shown by the horizontal lines in Fig.~\ref{fig:allowedwindow}, reflecting their $g_{h}$-independence. Since $T_{rec}$ in this regime cannot be lower than $T_{0}$, for the oscillation parameters suggested by the short-baseline hints flavor equilibration in this regime cannot be delayed below $T\sim 200$ keV.

\mysection{$T\sim M$}{.}
In fact, $T_{rec}$ cannot be lowered all the way to $T_{0}$. When $T\sim M$, there is an additional contribution to $\sigma_{h}$, from the $s$-channel $\nu\bar\nu$ diagram. To fully model this regime requires numerical treatment, which also takes account of the changing sign of the matter potential. Still, a useful and physically motivated result can be obtained by estimating the effect in the Weldon limit and then connecting the solution to Eq.~(\ref{eq:Treccontact}).

The peak unsuppressed $s$-channel cross section in the center of mass (CM) frame is $\sigma_{cm}\sim M^{-2}$, but not all neutrinos participate, only those for which the CM energy falls within the resonance width $\sim g^{2}M$ from the peak. The number density of such neutrinos is $n_{res}\sim T_{h} M \Gamma$ ({\it cf.} \cite{Friedland:2007vv}). Two other physical effects should be included: the relative velocity and the boost factor for the cross section. Both are parametrically important: for $T > M$, the two neutrinos in the resonant pair are moving nearly collinearly, hence $v_{rel}=|\vec{v}_{1}-\vec{v}_{2}|\sim\theta\sim M/T$; and the cross section for the pair is Lorentz-boosted \emph{perpendicular} to the c.o.m. collision direction, which yields another factor of $M/T$ \cite{Friedland:2007vv}. The rate $\Gamma_{fc}$ is $\Gamma_{fc}^{s-ch}\sim \sin^{2}\theta_{m} \sigma_{CM}(M/T)n_{res}v_{rel} \sim (\Delta m^{2}\sin2\theta)^{2} M^{2} g_{h}^{-2} T^{-5}$.

Requiring $\Gamma_{fc}^{s-ch}\sim T^{2}/M_{pl}$, we obtain
\begin{equation}
\label{eq:TrecmasslessSchannel}
T_{rec}^{s-ch} \sim (T_{0}^{5}M^{2})^{1/7} g_{h}^{-2/7}.
\end{equation}
Notice that the dependence on $g_{h}$ does not fully cancel in this regime. The constraint on the parameters of the model is obtained by requiring $T_{rec}^{s-ch}<T_{dec}^{e\nu}$:  
\begin{equation}
\label{eq:masslessSchannel}
g^{2}\gg M^{2} T_{0}^{5}/(T_{dec}^{e\nu})^{7}.
\end{equation}

The result applies so long as $T_{rec}^{s-ch} >M$. The power law Eq.~(\ref{eq:TrecmasslessSchannel}) intersects $T=M$ at 
\begin{equation}
\label{eq:Schannelcutoff}
M \sim T_{0} g_{h}^{-2/5}.
\end{equation}
For higher mediator masses, numerical treatment is required. However, as seen in Fig.~\ref{fig:Tdec}, the decoupling temperature quickly transitions to the NR regime of Eq.~(\ref{eq:Treccontact}). Eqs.~(\ref{eq:masslessSchannel}) and (\ref{eq:Schannelcutoff}) provide a good description of the triangular feature seen in Fig.~\ref{fig:allowedwindow} at 10 keV $\lesssim M \lesssim$ 100 MeV and $10^{-6}\lesssim g_{h} \lesssim 10^{-2}$.

Since $V_{m}$ changes sign at $T\sim M$, there is a narrow range of temperatures where $\theta_{m}\sim\pi/4$. One might worry appreciable production of $\nu_{h}$ occurs in this regime. However, as will be shown elsewhere~\cite{inprep}, this is not the case. 

\mysection{$l_{osc}>l_{col}$}{.}
Lastly, we consider the over-damped regime, which is responsible for low $T_{rec}$ on the left of Fig.~\ref{fig:Tdec}. We assume $T\gg M$, which will be justified {\it a posteriori}. The flavor conversion rate is $P(\nu_{a}\rightarrow\nu_{h})\Gamma_{fc}\sim\sin^{2} 2\theta (\Delta m^{2}/E)^{2}/\Gamma_{fc}\sim\sin^{2} 2\theta (\Delta m^{2}/T)^{2} M^{2} g^{-4} T^{-3}$. The equilibration condition, $\Gamma_{fc}\sim T^{2}/M_{pl}$, becomes
\begin{equation}
\label{eq:Tdamped}
T_{rec}^{damped} \sim T_{0}^{5/7} M^{2/7} g^{-4/7}.
\end{equation}

Thus, for light mediator masses $M<T_{0}$ and sufficiently large coupling $g$, equilibration could in fact occur below $T_{0}$. Yet, this is not a viable option, as it runs afoul of another cosmological constraint: Planck  requires \cite{Friedland:2007vv} that most neutrinos have to be free-streaming by the epoch of CMB formation, $T\sim 1$ eV. The criterion for free-streaming, assuming the mediator is heavier than 1 eV, is $\sigma n \sim \sin^{2}\theta g^{4} T^{5}/M^{4} < T^{2}/M_{pl}$ ({\it cf}. \cite{Friedland:2007vv})), or
\begin{equation}
\label{eq:Tstream}
T^{FS} \sim M (\sin^{2}\theta g^{4}M_{pl}/M)^{-1/3}.
\end{equation}
It is easy to verify that $T^{FS}> 1$ eV and $T_{rec}^{damped} < 1$ MeV cannot be simultaneously fulfilled. 

This argument reveals the physical reason why the neutrino portal framework is constrained to a window of parameters. The hidden interactions must delay recouping of flavors at $T\sim 1$ MeV, but at the same time, they should not  couple neutrinos at $T\sim 1$ eV.

\begin{figure}[tbh]
  \includegraphics[angle=0,width=0.47\textwidth]{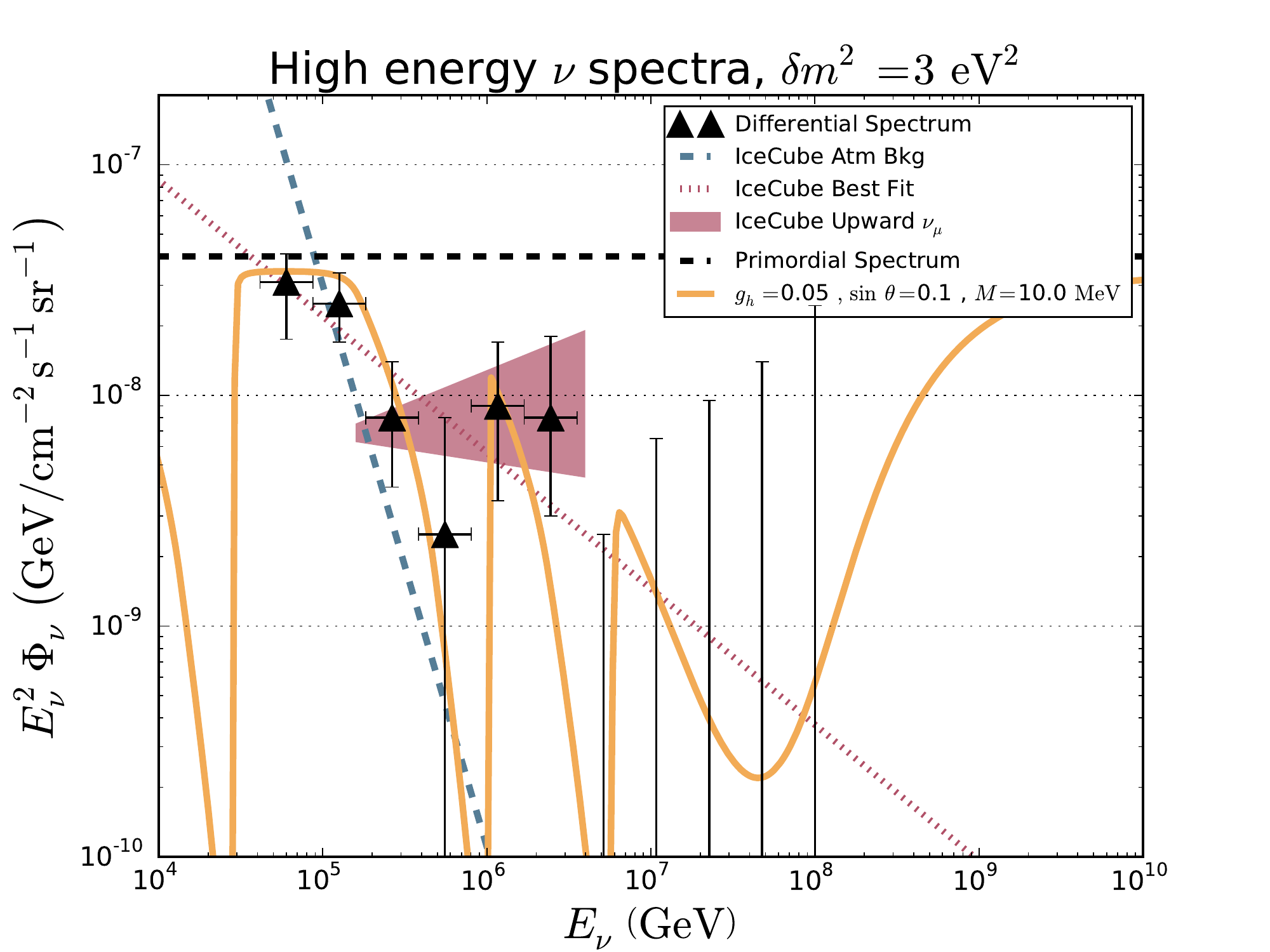}
  \caption{Illustration of the absorption features in the spectrum in the UHE astrophysical neutrinos~\cite{Aartsen:2015zva}.}
  \label{fig:Icecube}
\end{figure}

\mysection{IceCube}{.}
The next obvious step is to test neutrino-neutrino scattering in today's universe.  
The PeV astrophysical neutrinos recently detected at IceCube furnish a nearly perfect setup. These neutrinos travel to us from sources at cosmological distances, passing through the background of relic neutrinos. In the SM, these two populations do not see each other. In the presence of the hidden force, this conclusion is not automatic. The PeV neutrinos can scatter on the SM-like mass eigenstates, which contain an admixture of $\nu_{h}$, as well as on the mostly-$\nu_{h}$ component, which is also present thanks to the late-time $\nu_{a}\rightarrow\nu_{h}$ conversion.

For a relic neutrino with the mass $m_{atm}\sim 0.1$ eV, the CM energy of the interaction with a PeV neutrino is $\sqrt{2 E m_{atm}}\sim 10$ MeV, which falls right into our allowed window. This highly serendipitous fact makes for some unique observational signatures, as seen in Fig.~\ref{fig:Icecube}. The dips occur at several energies, corresponding to the resonant conditions for each of the neutrino mass eigenstates. With present statistics, it is not possible to conclusively confirm or rule out the existence of such features, though the hints in the data of a gap in the $0.4-1$ PeV range and a cutoff at 2 PeV are intriguing. The future 10 km$^{3}$ upgrade~\cite{Aartsen:2014njl} should furnish a conclusive test~\cite{inprep,Shoemaker:2015qul}. The sensitivity reach of this future measurement is shown in Fig.~\ref{fig:allowedwindow} by the red triangles, as will be explained in \cite{inprep}. 

Already with the present data, it is possible to exclude parameter values that predict so much absorption as to create a local horizon for neutrinos~\cite{Cherry:2014xra}. The part of this constraint that does not overlap with the free-streaming limit from Planck is indicated by \emph{IC horizon} in Fig~\ref{fig:allowedwindow}.

\mysection{Outlook}{.}
We have shown that the neutrino portal explanation to the oscillation anomalies is  constrained by the current data to a specific parameter window. Importantly, this window is testable in the near future. For example, Fermilab is presently developing a program aimed at conclusively testing these anomalies in the next 5-10 years~\cite{Antonello:2015lea}. Further tests will be carried out by reactor experiments, such as PROSPECT~\cite{Ashenfelter:2015uxt}. It is timely to ask what implications a positive result from these experiments would have for astrophysical and cosmological neutrinos. A deviation of $N_{eff}$ from its canonical value of 3.04, cosmological signatures of  neutrinos with eV mass, and absorption features in the upgraded IceCube detector could open a portal to a new sector of particle physics. 

\begin{acknowledgments}

AF was supported at SLAC by the DOE, Contract DE-AC02-76SF00515. IMS thanks  the Pennsylvania State University and the Institute for Gravitation and the Cosmos for support. JFC thanks the Center for Neutrino Physics at Virginia Tech for their support.
\end{acknowledgments}

\section{Appendix}

\mysection{Flavor equilibration mechanism}{.}
The populations of $\nu_{a}$ and $\nu_{h}$ can come into equilibrium via a combined action of oscillations and collisions. Oscillations are obviously necessary, as without them the flavor content of the system is conserved. Without collisions, however, equilibration is not achieved. All that happens to an initially active neutrino is oscillations according to the standard probability formula $P(\nu_{a}\rightarrow\nu_{h})=\sin^{2} 2\theta\sin^{2}(\Delta m^{2}L/4E)$, where $L$ is the distance, $E$ is neutrino energy, $\theta$ is the mixing angle and $\Delta m^{2}$ is the mass-squared splitting. For small $\theta$, the resulting $\nu_{h}$ content of the system would be small. 

The crucial property of collisions is their flavor-specific nature, which in our case is present by construction, since $\nu_{a}$ interact only via the SM weak interactions (WI), while $\nu_{h}$ interact only via the hidden force.  A scattering event is then in effect a measurement,  projecting the oscillated state onto the $|\nu_{a}\rangle$, $|\nu_{h}\rangle$ flavor basis and resetting the oscillation phase. The effective flavor change rate, $\Gamma_{fc}$, is a product of the oscillation probability between collisions $P(\nu_{a}\rightarrow\nu_{s})$ and the collision $\Gamma_{wi}$ rate:
\begin{equation}
\label{eq:convrateApp}
\Gamma_{fc}=P(\nu_{a}\rightarrow\nu_{s})\Gamma_{wi}=P(\nu_{a}\rightarrow\nu_{s})\sigma_{wi} n. 
\end{equation} 
The $\nu_{a}$ and $\nu_{h}$  equilibrate when $\Gamma_{fc}$ exceeds the expansion rate $\Gamma_{exp}\sim T^{2}/ M_{pl}$, where $M_{pl}$ is the Planck mass.

\mysection{Equilibration of the TS neutrinos}{.}
Let us consider in the limit $g_{h}\rightarrow 0$, which is the well-studied TS neutrino case. The WI collision rate, $\Gamma_{wi}=\sigma_{wi} n\sim G_{F}^{2} T^{2}  T^{3}$, rises rapidly -- while the typical oscillation rate, $\Delta m^{2}/T$, falls -- with $T$. The crossover point is
\begin{equation}
\label{eq:ToptimalApp}
 T_{opt}^{wi} \sim ( \Delta m^{2}/G_{F}^{2})^{1/6}.
\end{equation}
Here $G_{F}$ is the Fermi constant. For $\Delta m^{2}\sim 3$ eV$^{2}$, this gives $T_{opt}^{wi} \sim 50$ MeV. At this temperature, the frequencies of collisions and oscillations match, and the rate $\Gamma_{fc}$  is approximately maximized.
It is easily checked that this rate considerably exceeds the expansion rate $\Gamma_{exp}$, so that flavor is equilibrated at $T_{opt}^{wi}$. 

In fact, equilibration is achieved in a window of temperatures around $T_{opt}^{wi}$. Its upper end is dictated by the so-called ``Quantum Zeno'' effect. When $T>T_{opt}^{wi}$, collisions are more frequent than oscillations and flavor conversion is inefficient \cite{Stodolsky:1986dx}. The amount of $\nu_{s}$ generated between successive collisions is $P(\nu_{a}\rightarrow\nu_{s})\sim\sin^{2} 2\theta (\Delta m^{2}/T\Gamma_{wi})^{2}$ giving $\Gamma_{fc}=P(\nu_{a}\rightarrow\nu_{s})\Gamma_{wi}\sim\sin^{2} 2\theta (\Delta m^{2}/E)^{2}/\Gamma_{wi}\sim\sin^{2} 2\theta (\Delta m^{2})^{2} G_{F}^{-2}T^{-7}$, which rapidly \emph{decreases} with $T$.  Flavor conversion is out of equilibrium above
\begin{equation}
\label{eq:QZApp}
 T_{dec} ^{high}\sim (\sin^{2} 2\theta (\Delta m^{2})^{2} G_{F}^{-2} M_{pl})^{1/9}.
\end{equation}

When $T<T_{opt}^{wi}$, collisions are less frequent than oscillations and, on average, $\bar{P}(\nu_{a}=\nu_{s})=(1/2)\sin^{2} 2\theta$. Equating $\Gamma_{fc}$ and $\Gamma_{exp}$ gives
\begin{equation}
\label{eq:TtsApp}
 T_{dec}^{low} \sim [G_{F}^{2} M_{pl}(1/2)\sin^{2} 2\theta]^{-1/3}.
\end{equation}
The lower end of the window is due to the freeze-out of collisions. 
Indeed, the SM process $\nu\bar\nu\leftrightarrow e^{-}e^{+}$  freezes out at
\begin{equation}
\label{eq:TenuApp}
 T_{dec}^{e\nu} \sim (G_{F}^{2} M_{pl})^{-1/3} \sim 1\mbox{ MeV},
\end{equation}
which is found by equating $\Gamma_{exp}$ and the WI collision rate, $G_{F}^{2} T^{5}$ \cite{Kolb:1990vq}. Up to the factor of the mixing angle, the two are the same. This is the root cause of difficulties for the TS explanation of the short-baseline anomalies: since the oscillation fit requires $\theta$ to be not too small, the flavor equilibrium will be achieved slightly above $T_{dec}^{e\nu}$.

Two comments are in order. First, the $\nu\bar\nu\leftrightarrow e^{-}e^{+}$ freezeout is not instantaneous and happens over a range of temperatures, $\sim1-3$ MeV. For our conservative estimate in Fig.~\ref{fig:allowedwindow}, we took $T_{dec}\sim3$ MeV, but $T_{dec}\sim1$ MeV, or even 0.5 MeV, still yields a fractional deviation of $N_{eff}$ from 3.04 that could be observable in future.  

Second, we omitted the effects of plasma on neutrino oscillations. It is standard knowledge \cite{Raffelt:1996wa} that the WI generate a second-order Notzold-Raffelt matter potential \cite{Notzold:1987ik} $V_{NR}=-(8\sqrt{2}/3)G_{F} E [ (n_{\nu}\langle E_{\nu}\rangle  +n_{\bar\nu}\langle E_{\bar\nu}\rangle  )/m_{Z}^{2} + (n_{e^{-}}\langle E_{e^{-}}\rangle  +n_{e^{+}}\langle E_{e^{+}}\rangle  )/m_{W}^{2}]$. (The first-order  $\sqrt{2}G_{F}n$ terms largely cancel between matter and antimatter.) This potential, however, does not change the conclusions. Indeed, on the lower end of the window, Eq.~(\ref{eq:Tts}), $V_{NR}\sim G_{F}^{2}T^{5} \ll \Delta m^{2}/T$, so that the oscillation parameters are not modified. 
\footnote{On the upper end, the oscillations parameters are modified, $\Delta m^{2} \rightarrow V_{NR}$ and $\sin 2\theta \rightarrow \sin 2\theta (\Delta m^{2}/V_{NR} T)$, but their product entering Eq.~(\ref{eq:QZ}) remains approximately the same.}

\begin{figure}[htb]
  \includegraphics[angle=0,width=0.47\textwidth]{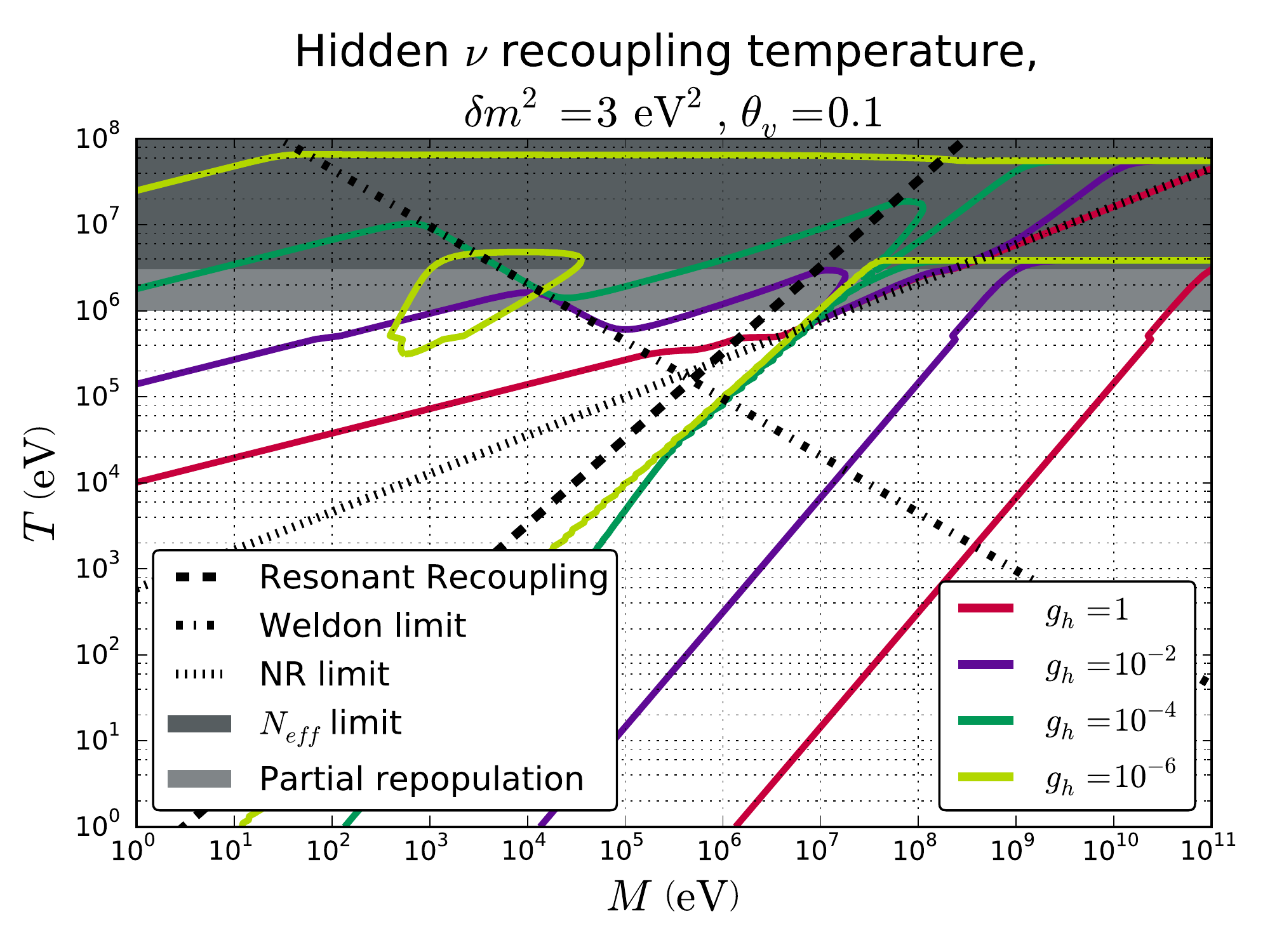}
  \caption{The dependence of the recoupling temperature on the mediator mass $M$ for several values of the coupling $g_{h}$. The gray regions have the same meaning as in Fig.~\ref{fig:allowedwindow}.}
  \label{fig:Tdec}
\end{figure}

\mysection{}{}
In Fig.~\ref{fig:Tdec}, we plot the equilibration temperature as a function of the mediator mass $M$ for several values of the coupling $g_{h}$.

\bibliography{recoupling}

\end{document}